\documentstyle[twocolumn,aps]{revtex}
\begin{document}

%
%

\def\be{\begin{equation}}
\def\ee{\end{equation}}
\def\bearr{\begin{eqnarray}}
\def\eearr{\end{eqnarray}}
\def\tc{$T_c~$}
\def\tcl{$T_c^{1*}~$}
\def\c2{$ CuO_2~$}
\def\lsco{LSCO~}
\def\bi{Bi-2201~}
\def\tl{Tl-2201~}
\def\hg{Hg-1201~}

\title{ Competition between Superconductivity and Charge Stripe Order in
High \tc  Cuprates }

\author{G. Baskaran\cite{email}}

\address{ The Institute of Mathematical Sciences,\\
                      Madras 600 113, India\\
and\\
Institute for Theoretical Physics\\
University of California\\
Santa Barbara, CA 93106-4030, USA}

\maketitle

\begin{abstract}

Members of the high \tc curapte family (with defect free 
\c2 planes) are suggested to have an `{\em Intrinsic single
layer superconducting} \tc', larger than the experimentally
observed \tc, at a given hole concentration.
This difference occurs to varying degrees among the different
members, due to an anomalous response of the d-wave superconducting
state to self generated charge perturbations. In particular, 
any quasi elastic charge stripe order, arising from electron-electron 
and electron- lattice coupling, can suppress 
the large intrinsic superconducting \tc. Existing experimental 
results on the wide \tc  variation in the single layer cuprate 
families, such as the low \tc ($\approx 38 K, 33 K$) of \lsco and \bi
compared to high \tc ($\approx 95 K, 98 K$) of \tl and \hg, 
the strain induced increase in \tc (25 to 50 K) observed in 
\lsco are qualitatively explained 
by our mechanism. We predict that under proper epitaxial strain
the \tc of \lsco and \bi should increase from 30's all the way
upto the intrinsic \tc  $\sim $ 90 K, at optimal doping..   

\end{abstract}

\section{Introduction}

The superconducting \tc of cuprates$^{1,2}$
 has a very wide variation 
from nearly zero all the way up to 165 K among the family members 
at optimal doping.  With improved experiments it is 
becoming clear that most of these variations are intrinsic and
are not driven by lattice defects or dopant induced disorder.
From the materials science point of view, a thorough understanding
of this phenomenon may provide suggestions for reaching even higher
transition temperatures; from basic physics point of view this 
should help us sharpen our theoretical insights on the mechanism
of cuprate superconductivity.

Strong electron correlations in the copper-oxygen based narrow
d-band that is believed$^3$ to produces superconductivity 
is also capable of producing charge, spin and lattice related 
instabilities.  In fact, charge stripe order seems generic to 
perovskite structures; any soft, rotational and distortional modes
of the octahedra can enhance the charge localization
(ordering) tendency.  In particular, recent 
experiments$^{4-8}$
 and theories$^{9-11}$
 provide growing evidence for certain 
low energy dynamical charge ordering tendency called stripes.
A natural question is {\em Do the charge stripe correlations, when
they are dynamical with low frequency, provide a new mechanism for 
superconductivity or do they 
help or fight against an existing superconductivity mechanism}.
A recent numerical analysis by White and Scalapino$^{11}$
already shows that charge stripes tend to suppress pairing correlations.
Various mean field theories also bring out the presence of charge
stripe correlations at low doping, before the emergence of 
superconductivity.

The aim of the present paper is to i) propose an `Intrinsic single 
$CuO_2$ layer superconducting \tc ' hypothesis,  
ii) discuss the physics behind the charge stripe formation and 
iii) suggest that an anomalous response of the d-wave state to
low energy or quasi static charge stripe fluctuations results in a 
large decrease of the intrinsic one-layer transition temperature \tcl.  

We base our suggestions on a comparative study of the existing 
experimental results on the one layer family of 
cuprates, some existing numerical results and our own theoretical 
considerations.  Our major emphasis is the observation that 
within the one layer family the superconducting $T_c$ varies widely 
from 5 K to 98 K, even though the normal state remains reasonably 
unchanged.  While impurities and doping induced disorder can cause 
part of this variation, we will argue that it is an anomalous 
response of the d-wave state to the development of 
low energy dynamical charge stripes,  that causes the 
large suppression of an intrinsic superconducting \tcl of a single 
$CuO_2$ layer. 

Some of the existing experimental results on the anomalous increase
of \tc of \lsco under epitaxial strain as well as some pressure 
dependence of \tc can get qualitatively explained by our mechanism.
One of our predictions is that under a proper epitaxial strain
the \tc of optimal doped \lsco  as well as \bi should increase all
the way upto 95 K.

\section{Intrinsic single plane \tc hypothesis and experimental results}

We hypothesize that an ideal rigid (with immobile atoms) 
single layer of \c2 has an 
intrinsic superconducting transition temperature \tcl for a given hole
density x.  The experimentally observed large variation
in \tc among the families of cuprates must mean, by 
the above hypothesis, that there is some other mechanism
or interaction that is either decreasing or increasing the 
intrinsic \tc at a given doping.  From theory point of view 
an idea of the intrinsic \tc and its $x$-dependence can be 
obtained from the early RVB mean field theory of 
superconductivity$^{12}$ and their modern refined  
versions$^{13,14}$

The maximum \tc of single layer materials \lsco, \bi, \tl and
\hg are 38, 33, 95 and 98 K respectively.  In spite of the 
varying \tc most of the ab plane properties including 
the T dependent resistivity $\rho_{ab}$ at optimal
doping are nearly the same in all the four materials.
A major difference that distinguishes the low \tc \lsco and \bi
from the high \tc \tl and \hg are the charge stripe tendencies
in the former below the spin gap temperature scale.  
Neutron$^4$, X-ray$^5$ and NQR$^6$
studies show strong charge
ordering tendency in \lsco.  Hall angle studies in the ab-plane
of \lsco and \bi show$^{8,15}$
 a temperature dependence of the form $T^{2-\alpha}$
(with $\alpha \approx 0.3$) as opposed to $T^2$ in \tl$^{17}$
 and most likely in \hg.  The $T^{2-\alpha}$ dependence is easily  
interpreted as arising from the scattering of spinons by the
incipient dynamical stripes$^{16}$.

We assume that the \tc of \tl is closest to the intrinsic
\tcl.  This is because experiments indicate that
there is very little inter layer pair tunneling contribution to the 
superconducting condensation energy, suggesting that the
mechanism of superconductivity that is operative is dominantly
a single \c2 layer phenomenon. Thus we identify the maximum
\tc $\approx 95$ with the intrinsic \tcl$= 95$ at optimal
doping. 

Another experimental input that we use and also explain is
the doubling of critical temperature from 25 K to 50 K
in $La_{1.9}Sr_{0.1}CuO_4$ under compressional epitaxial strain
observed by Locquet et al.$^{18}$. 

Numerical experiments on large U Hubbard model have brought out 
enhanced pair susceptibility with $d_{x^2 - y^2}$ symmetry 
as well as charge stripe susceptibility.  The work by White and 
Scalapino$^{11}$ indeed shows how in the presence of $t'$ 
term the stripes
are suppressed at the expense of superconductivity.  The wave vector at 
which the charge susceptibility is largest in numerical studies$^{19}$
is close to the experimentally seen wave vectors, $ 4\pi ( x, 0)$ 
and $ 4\pi (0,  x) $, where $x$ is the hole density.

\section{mechanism of charge order}

Central to stripe order at low doping is the interplay between 
one electron kinetic energy (t)  and the super exchange energy (J), 
and to some extent short range repusion energy. 
In the conducting state the super exchange term
favours short range singlet bond formation and hole delocalization 
frustrates this tendency and vice versa. In the cuprates 
the parameters t and J have comparable values of $\sim 0.2~{\rm and}~
0.15 {\rm eV}$.  The doped `holes' have an option
to get delocalized isotropically in the plane and remain less  
coherent - some what like a collection of Brinkman-Rice holes
with enhanced short time scale spin scrambling.
This state will pave the way for the d-wave superconducting state
at low temperatures.
The other option is to delocalize slightly {\rm preferentially} 
along a subset of parallel chains and become more coherent at least in one
direction - this will be the charge stripe state. In the charge stripe
state the enhanced quasi one dimensional motion of holes reduces the
spin scrambling. 

Very recent ARPES$^7$ brings out the high frequency 
(corresponding
to a scale of about 0.5 eV) character of the stripe fluctuations.
The stripe correlations quantum melt as we come to low frequencies
and eventually disappear, leaving way for superconductivity. 
In some compounds perhaps it co-exists with superconductivity.
We suggest that the stripe order that show up in mean field
theories$^{20}$ should disappear or be strongly suppressed
with proper inclusion of quantum fluctuations.  

From many body theory point of view the high frequency insulating
stripe correlations are analogue of {\em the strong short range crystalline
order in liquid $He^3$ below the melting line at very low temperatures}.
In liquid $He^3$ the crystalline correlation does not have a quasi 
one dimensional character.  From the point of view of gaining
delocalization energy the `crystalline correlations' in the \c2 planes
have quasi one dimensional character. These `crystalline 
correlations' are the Mott insulating stripes that separate the more 
conducting `chains'.  In a sense the insulating stripes are short 
time scale memory of the parent Mott insulating state.  

From the point of view of gain in kinetic energy 
the conducting stripe should be nearly `quarter' filled,
as also seen experimentally.  The reason for this is that 
the holons along a conducting chain maximise their kinetic energy
when they form a half filled band. This is easily seen in the limit
$J \rightarrow 0$ where the spinless fermions maximise their
 kinetic energy gain in a chain only at half filling.  
Notice that in the one dimensional t-J model a quartered filled band 
of electrons corresponds to a half filled band of holons.  Thus the 
major factor that
determines the ordering wave vector is the mean distance 
between the nearly `quarter' filled charged stripes at a given doping.
Under this assumption the ordering wave vectors become
$4\pi(x,0)~{\rm and}~4\pi(0,x)$.   

Another important point we wish to make is that the high 
frequency stripe fluctuation can be frozen out if there is sufficient
help from phonons.  This is already visible in various numerical
results where the boundary conditions stabilize stripes in the 
ground state. If we have a sufficiently soft phonon corresponding
to the ordering wave vector and also strong coupling of the charge
density fluctuations to this phonon, the lattice
distortion and the stripe order can support each other 
self consistantly leading to quasi static or even true long range order. 

We have done a simple RPA calculation coupling the hole density
fluctuations to phonon and find that for the realistic parameters
the freezing of the stripe order can occur.  Our modelling involves 
coupling of the holon density of the spin gap phase to phonons.
Assuming a model dispersion for the holons, the RPA  
expression for phonon frequency$\hbar \omega_Q$  at the nesting 
wave vector $Q$ is given by the self consistent equation:

\be
\hbar \omega_{Q} \approx \hbar \omega_0 - {\lambda^2 \over N }
\sum {{1} \over { \hbar \omega_Q + 2 \hbar v_{c} k } }   
\ee

where $\hbar\omega_0$ is the bare phonon frequency, $\lambda$
is the holon-phonon coupling constant and $v_c$ is the holon
velocity close to the chemical potential. 
The second term in the above expression is the particle-hole
susceptibility for the holons.  In the spin gap phase
the holons do not have an extended fermi surface, and hence  
the particle-hole susceptibility does not lead to 
a logarithmic divergence. However, the smallness of the 
holon velocity increases the pair susceptibility and equation (3)  
exhibits a zero frequency solution, and the onset of charge 
stripe instability, at a non zero critical value of 
$\lambda_c \sim \sqrt{ \hbar \omega_0 \epsilon_F { v_c \over v_F} }$.  
This large value is only an upper bound, as we have not
introduced important short range repulsion among the quasi particles 
that further encourage density wave instabilities.  A modification
of the susceptibility in the equation for the phonon frequency
(equation 1) by a renormalized susceptibility 
$ \chi(q.\omega) \rightarrow { { \chi(q,\omega)}\over
{ 1 - U \chi(q,\omega)}}$ shows large reduction of the critical
coupling constant.  Here U is an effective short range repulsion
among the quasi particles.  

\section{How does charge order instability suppress  
superconducting \tc ?}

In many body systems, when two or more instabilities occur simultaneously, 
they need not support each other. There may be some repulsive 
couplings between the two orders that will supress one in the presence 
of the other. This is a kind of level repulsion
between the two different broken symmetry vacuua. 

In this section we bring out a mechanism that brings out this repulsion
quantitatively.  It is well known that the response of d-wave superconducting 
state to static disorder is strong and anomalous, analogous to the
response of s-wave superconductor magnetic impurities. Zn doping 
in cuprates illustrates this well. This makes the superconducting 
\tc very sensitive to development of low frequency periodic or random 
modulation of charges.

In the case of d-wave, unlike the s-wave, a finite density of 
scatterers leads to a finite quasi particle density of 
states$^{21,22}$ at the chemical potential.  
These low energy states 
emerge by depleting the d-wave condensate through a  
quantum interference effect arising from the changing sign
of the d-wave amplitude in k-space. An approximate 
expression$^{23,24}$ for the resulting reduction in \tc is
\be
k_B \Delta T_c \approx {n_d \over {4N(E_F)}} \sin^2(\delta)
\ee
where $n_d$ is the density of scatterers and $\delta$ is the 
phase shift arising from the individual scatterers. It is 
remarkable that the reduction in \tc is universal in the sense
it is independent of the bulk \tc ! 

We use this formula to understand the sensitivity of 
\tc of the \c2 layer to the development of low frequency 
charge stripe correlations. {\em The low frequency charge stripe  
correlation provides a self consistent `random'
potential for the electrons}.  The above formula is applicable 
as long as the charge stripe fluctuation frequency 
$ \nu << {\Delta(0)\over \hbar}$.  The typical gap value 
$\Delta(0)$ is of the order of 20 meV in cuprates and the 
charge stripe correlation that has been seen by NQR, 
X-ray and neutron scattering are observed at much lower 
frequencies.  Because of its dynamical character, as well
as the non fermi liquid character of the normal state, the 
incipient holon density fluctuations affect the normal 
state transport less dramatically than they do \tc.

We can use the above formula to get an estimate of the 
reduction in \tc, and substitute in equation (2) for 
$n_d$ the rms amplitude of the charge stripe order parameter, and
for $\delta$ the phase shift experienced by the rest of 
the electrons due to the presence of a quasi static 
charge order over a length scale of the order of the 
coherence length of the cooper pair.

If we assume a phase shift of $\pi\over2$ , an rms 
amplitude of 1\%  leads to a reduction in \tc of about 
50 K. The experimentally observed \tc 
in \lsco as well as \bi at optimal doping 
is reduced from our definition of intrinsic value of 95 K
by about $ 60 K$.
The above is a very rough estimate, but tells
us that \tc can get reduced considerably.

From experiments we can infer that the charge stripe order
can not bring the \tc lower than 33 K at optimal doping, and
perhas superconductivity can co-exist with charge stripe
order at these low temperatures.  

Our conclusion can be introduced as a phenomenological 
term in the Ginzburg Landau free energy  as a coupling 
between the charge stripe order parameter $\phi({\bf r})$
and the d-wave order parameter $\psi ({\bf r})$:
\be
 H_{int} \approx  g \int  d{\bf r} \phi^2({\bf r}) 
\psi^*({\bf r})\psi({\bf r}) 
\ee
Since $g \approx {{\sin^2 \delta}\over {4N(E_F)}}$ 
is positive the charge stripe fluctuation always decreases \tc.

\section{Why do LSCO and \bi yield to charge stripe  
instability in contrast to \tl ?}

We suggest that the enhanced charge stripe tendency (seen 
experimentally) in LSCO and Bi-2201 is responsible for their 
low transition temperatures. This means that phonons in
LSCO and \bi corresponding to the nesting wave vector are 
softer and strongly coupled to the
electrons in comparison with Tl-2201.  We can justify this statement
using a recent experiment on LSCO and in the process also explain
the experimental result itself. 

Locquet et al.$^{18}$ have found a remarkable result 
that under a compressive 
epitaxial strain the superconducting Tc of $ La_{1.9} Sr_{0.1} Cu O_4 $ 
is increased from 25 K to 50 K, while the ab plane resistivity remains
unchanged above the spin gap phase temperature. We argue that the compressive 
epitaxial strain hardens the octahedral rotational and distortional phonon 
mode corresponding to the 
stripe order.  In the process the charge stripe instability is removed and
most of the latent superconducting order brought back.

Our explanation of the anomaly observed Locquet et al. also gives
us a clue as to why Tl-2201 escapes the charge stripe instability.  A closer 
look at the structure of Tl-2201 reveals that the 2d octahedral
layer of $CuO_4$ is sandwiched between two layers of $ (Tl O)_2$.
The natural lattice parameter of the $(Tl O)_2$ is likely to be
different from that of the 2d octahedral layer of $CuO_2$.  Thus
the $CuO_2$ layer is either under a compressive strain or 
expansional strain provided by the $(TlO)_2$ layers.  This should
harden the octahedral rotational and distortional phonon mode 
corresponding to the
stripe order, there by escaping the charge stripe instability.  Indeed, the 
ab plane lattice parameter of Tl-2201 and LSCO are 3.864 Au and
3.787 Au respectively, indicating the relative stretch and the
consequent stiffness of the \c2 layer in Tl-2201.

In the same fashion in the one layer \hg dumb bell co-ordination
of Hg and the intervening oxygen of the $HgO$ plane seems to stiffen
the \c2 planes there by discouraging charge stripe order.
In \hg in the presence of $Hg$ deficiency one has already 
seen$^{25}$ evidence of stripe tendency, in the 
sense of decrease of \tc  around the magic hole density of $x = 
{1\over8}$. This result can be interpretted as the beginning of 
the stripe tendency arising from $Hg$ deficiency and the consequent
lattice softening.

Our observation also qualitatively explains the mystery of the strong 
increase in \tc of oxygenated LSCO in pure state as well as in thin 
film epitaxy observed$^{26}$ in many experiments.  
Several intriguing pressure dependences of \tc$^{27}$ 
 in a family of cuprates can be perhaps explained by our 
mechanism. 

\section{conclusions}

In general we can say that {\em any lattice strain
arising from sandwiching layers or from 
ionic radii mismatch of cation,  that hardens the lattice mode
corresponding to the charge stripe ordering wave vector, will disfavor
charge stripe order and encourage superconductivity.} This is 
consistent with the finding of Attfield  et al.$^{28}$
 where they find 
a systematic increase of the optimal \tc with cation radius.

Based on our theory we also predict that one  
can increase superconducting Tc of \lsco and \bi from 
35 K to nearly 95 K by producing appropriate epitaxial strain 
at optimal doping.

Ong$^{29}$ also suspects exclusion of stripe fluctuations in the 
superconducting state that are stabilized by epitaxial strains 
in some of the samples studied by transport measurements.
It will be important to
study systematically the correlation between the development of
stripe order and the reduction in \tc as predicted by us,
in \tl, \hg and \bi systems.

We will discuss the case of bilayer and tri layer in a separate 
paper$^{30}$, where there are interesting competition between the 
single layer superconductivity, interlayer pair tunneling and
the charge stripe order.

\section{Acknowledgment}

I thank P.W. Anderson, N.P. Ong, Y. Ando, and V.N. Muthukumar for 
discussions; P.W. Anderson for hospitality at Princeton;
N.P. Ong especially for valuable experimental 
informations including references 18 and 26. This research was
supported in part by the National Science Foundation under 
Grant No. PHY9407194 and DMR-9104873.

\end{document}